\documentclass[
	aps,
    prr,
    superscriptaddress,
    twocolumn,
    a4paper,
    floatfix,
    longbibliography,
    raggedbottom,
]{revtex4-1}

\usepackage{amsmath}
\usepackage{amssymb}
\usepackage{bbm}
\usepackage{mathtools}
\usepackage{dsfont}
\usepackage{braket}

\usepackage{graphicx}
\usepackage[svgnames]{xcolor}

\usepackage{siunitx} 

\usepackage{ragged2e}
\usepackage{array}
\usepackage{tabularx}
\usepackage{booktabs}

\definecolor{cset-aps-blue}{RGB}{18,84,168}
\definecolor{cset-aps-limegreen}{RGB}{153,204,51}

\definecolor{cset-aps-blueberry}{RGB}{28,128,158}
\definecolor{cset-aps-turquoise}{RGB}{0,67,88}
\definecolor{cset-aps-limegreen}{RGB}{190,219,67}
\definecolor{cset-aps-darkblue}{RGB}{31,138,112}
\definecolor{cset-aps-yellow}{RGB}{255,225,25}
\definecolor{cset-aps-orange}{RGB}{253,116,0}
\definecolor{cset-aps-red}{RGB}{219,0,43}

\newcommand{\myVerticalCenter}[1]{\begingroup
\setbox0=\hbox{#1}%
\parbox{\wd0}{\box0}\endgroup}

\newcommand\myIconCube[1]{{\color{#1}\rule{1.5ex}{1.5ex}}}
\newcommand{\myMarker}[1]{(\myVerticalCenter{\myIconCube{#1}})}
\usepackage{tikz}

\usepackage{pgfplots}
\pgfplotsset{%
    every axis legend/.append style={%
        cells={anchor=west},
        at={(0.96,0.04)},
        anchor=south east,
        font=\scriptsize,
        },
    every axis/.append style={%
        yticklabel style={%
            /pgf/number format/fixed zerofill,
            /pgf/number format/precision=2},
        },
    width= \textwidth,
    height=8cm,
    xmajorgrids=true,
    xminorgrids=false,
    minor x tick num=1,
}
\usepgfplotslibrary{external}

\usetikzlibrary{decorations}

\newcommand{\D}[0]{\mathrm{d}}
\newcommand{\branch}[0]{{(\alpha)}}
\newcommand{\I}[0]{{\mathrm{i}}}
\newcommand{\e}[0]{{\mathrm{e}}}
\newcommand{\Eq}[1]{Eq.~\eqref{#1}} 
\newcommand{\Fig}[1]{Fig.~\ref{#1}} 
\newcommand{\state}{{\pm}}
\newcommand{\excited}{+}
\newcommand{\ground}{-}

\usepackage{hyperref}
\hypersetup{%
    colorlinks=true,
    linkcolor={cset-aps-red},
    linkbordercolor={cset-aps-red},
    filecolor={cset-aps-orange},
    filebordercolor={cset-aps-orange},
    citecolor={cset-aps-blue},
    citebordercolor={cset-aps-blue},
    urlcolor={cset-aps-blue},
    urlbordercolor={cset-aps-blue},
    menucolor={cset-aps-limegreen},
    menubordercolor={cset-aps-limegreen},
    breaklinks=true,
    pdfborderstyle={/S/U/W 2},
    pdfpagemode=UseOutlines,
    pdfstartpage={1},
}

\newcommand{\affDLR}{\address{Institute of Quantum Technologies, German Aerospace Center (DLR), S\"{o}flinger Stra\ss e 100, D-89077 Ulm, Germany}}
\newcommand{\affHAN}{\address{Institut f{\"u}r Quantenoptik, Leibniz Universit{\"a}t Hannover,
    Welfengarten 1, D-30167 Hannover, Germany}}
\newcommand{\affULM}{\address{Institut f{\"u}r Quantenphysik and Center for Integrated Quantum
    Science and Technology (IQ\textsuperscript{ST}), Universit{\"a}t Ulm, Albert-Einstein-Allee 11, D-89069 Ulm, Germany}}
\newcommand{\affSCHLEICH}{\address{Hagler Institute for Advanced Study and Department of Physics and Astronomy, Institute for Quantum Science and Engineering (IQSE), Texas A{\&}M AgriLife Research, Texas A{\&}M University, College Station, Texas 77843-4242, USA}}

\newcommand{\affSchubertCurrent}{\address{Current address:  German Aerospace Center (DLR), Institute for Satellite Geodesy and Inertial Sensing, c/o Leibniz Universit{\"a}t Hannover, DLR-SI, Callinstraße 36, D-30167 Hannover, Germany}}

\begin{document}

\title[Title]{Atom-interferometric test of the universality of gravitational redshift and free fall}
\collaboration{Published as
    \href{https://journals.aps.org/prresearch/abstract/10.1103/PhysRevResearch.2.043240}
    {Physical Review Research \textbf{2}, 043240 (2020)}}

\author{Christian Ufrecht}
\email{christian.ufrecht@gmx.de} 
\affULM

\author{Fabio Di Pumpo}
\affULM

\author{Alexander Friedrich}
\affULM
    
\author{Albert Roura}
\affDLR
    
\author{Christian Schubert}
\affHAN
\affSchubertCurrent
    
\author{Dennis Schlippert}
\affHAN

\author{Ernst M. Rasel}
\affHAN
    
\author{Wolfgang P. Schleich}
\affULM
\affDLR
\affSCHLEICH
    
\author{Enno Giese}
\affULM
\affHAN

\begin{abstract}
Introducing internal-state transitions simultaneously on each branch of a light-pulse atom interferometer, we propose a scheme that is concurrently sensitive to both violations of the universality of free fall and gravitational redshift, two premises of general relativity. In contrast to redshift tests with quantum clocks, a superposition of internal states is not necessary but merely transitions between them, leading to a generalized concept of clocks in this context. The experimental realization seems feasible with already demonstrated techniques.
\end{abstract}

\maketitle

\section{Introduction}

The phenomenal advance in accuracy of atomic light-pulse interferometers over the last decades has not only led to high-precision applications in gravimetry~\cite{DroppingAtoms,Syrte} and gradiometry~\cite{Gradiometrie1,Gradiometrie4}, but also allows for probes of fundamental physics such as through measurements of the fine-structure constant to constrain Standard-Model extensions~\cite{FineStructure1,FineStructure2,FineStructure4}, gravitational wave detection~\cite{GravitationalWaves} or tests of the universality of free fall~\cite{WEP1,WEP3,WEP4,WEP5,KasevichWEP}. Since the universality of free fall (UFF) and the universality of gravitational redshift (UGR) form the foundations of general relativity, their violation would directly hint towards new unknown physics.
While the former has been tested with light-pulse atom interferometers for two different atomic species to the $10^{-12}$ level~\cite{KasevichWEP}, interferometry based on clocks as input \cite{Zych1} is insensitive to UGR violations \cite{TwinPaper} as parametrized below.
In this article we propose an interferometer scheme sensitive to both violations of UFF and UGR (depicted in \Fig{fig:Interferometer}).
Whereas redshift sensitivity may arise from the initialization of a quantum clock during the interferometer sequence~\cite{RouraRedShift}, we show that a superposition of internal states that constitutes a clock is not necessary. Instead, the sensitivity originates solely from the interferometer geometry and change of internal states.

\section{Relativistic effects in atom interferometers}
In general relativity an ideal clock moving along a worldline $x^\mu$ measures proper time
\begin{equation}
 \tau=\frac{1}{c}\int \sqrt{\D x_\mu \D x^\mu}
\end{equation}
where $c$ is the speed of light.
This quantity is connected to the phase of a sufficiently localized matter wave via the relation~\cite{Hogan,RouraRedShift}
\begin{equation}
\label{eq:PhaseMatterWave}
\phi=-\omega\tau+S_\mathrm{em}/\hbar
\end{equation}
where $\omega=mc^2/\hbar$ denotes the Compton frequency, $m$ the mass of the atom, and $S_\mathrm{em}$ accounts for interactions with electromagnetic fields guiding the matter wave. 
Because the phase depends on proper time, it is conceivable that atom interferometers provide a platform for tests of special and general relativity.
We focus on UFF and UGR and propose a quantum-mechanical experiment testing both principles. 
For that, we define UFF and UGR tests operationally and in complete accordance with their classical counterparts.
UFF states that gravitational acceleration in a linear gravitational field is universal, while UGR assumes proper-time differences measured by two clocks in a gravitational field to be independent of the composition of the clocks.
Introducing violation parameters in Sec.~\ref{Violation model} will help us unambiguously identify violations of both principles in the interferometer phase through a comparison to predictions of textbook experiments.

In a light-pulse interferometer a series of short light pulses drives the atoms into a coherent spatially delocalized superposition and subsequently directs them along two branches of the interferometer.
Upon recombination, the relative phase, which according to \Eq{eq:PhaseMatterWave} depends on the proper-time difference $\Delta \tau$ between the branches, is inferred from the interference.

Since $\Delta \tau$ enters the phase proportional to the Compton frequency $\omega$, it was suggested \cite{MuellerRedShift} that, due to its immense value, light-pulse interferometers could push current bounds on violation parameters of UGR by several orders of magnitude.
However, standard closed light-pulse interferometers without internal transitions are insensitive to the gravitational redshift in a uniform field~\cite{RedshiftDebate2, RedshiftSchleich,RouraRedShift,SMI,TwinPaper,DiPumpo2020}, not showing any dependence on UGR violations as defined in Sec.~\ref{Violation model} because of two fundamental aspects: First, the proper-time difference between the branches is independent of gravity and consequently of purely special-relativistic origin \cite{TwinPaper}. Second, for a single internal state only one energy scale of order of the Compton frequency without any energy reference exists~\cite{RedshiftDebate1}.

To introduce the latter, one can envision an experiment where the atoms enter the interferometer in a localized internal superposition (constituting an atomic clock) and are then set into a coherent superposition of two branches by laser pulses~\cite{QuantumClockInterferometry}.
As a manifestation of Einstein's relation $E=mc^2$, a two-level atom of energy spacing $\hbar \Omega$ with mass $m_\ground$ in the ground state has a different mass $m_\excited=m_\ground+\hbar \Omega/c^2$ after excitation~\cite{Zych1, Sonnleitner,Schwartz}.
Thus, the Compton frequency  becomes state dependent, i.\,e. $\omega_\state=m_\state c^2/\hbar$.
In such quantum-clock interferometers the proper-time difference associated with the interferometer branches leads to a beating in the interference signal~\cite{Zych1}.
Contrary to symmetric schemes like Mach-Zehnder geometries or the sequence proposed in the present article, the proper-time difference of an asymmetric Ramsey-Bord\'{e} interferometer is non-vanishing, but of purely special relativistic origin \cite{TwinPaper}.
\begin{figure}
	\begin{center}
		\includegraphics{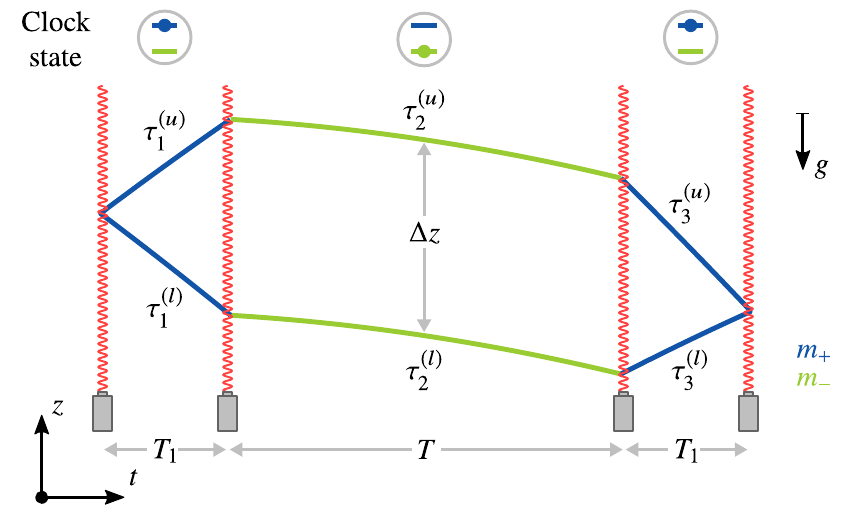}
		\caption{\textit{Redshift-sensitive geometry.} The interferometer is based on a diffraction scheme where each laser pulse, apart from transferring momentum, changes the internal state. Initially the input wave packet is split into two components by a $\pi/2$ pulse, which move in opposite directions with momentum $\pm\hbar k$, where $k$ is the effective wave number of the laser fields. At the second and third laser pulse the wave packets are redirected and finally recombined at the fourth laser pulse by further $\pi/2$ pulses. If initially in the ground state $m_\ground$ \myMarker{cset-aps-limegreen}, the atoms reside in the excited state $m_\excited$ \myMarker{cset-aps-blue} during the first and third segment but occupy the ground state during the central segment of the interferometer. The atoms are always in the same internal state at equal times (in the laboratory frame).
		} 
		\label{fig:Interferometer}
	\end{center}
\end{figure}
In contrast, if the internal superposition is created \emph{within} instead of \emph{before} the interferometer sequence, corresponding to an initialization of the clock at a variable time, the phase shift becomes sensitive to the gravitational redshift~\cite{RouraRedShift}.

We now show that a superposition of internal states is not necessary.
Instead, the energy reference mandatory for redshift tests is introduced solely by a sequence of \textit{internal transitions} at different times of the interferometer
so that the atoms occupy the same internal state at equal times (in the laboratory frame), see \Fig{fig:Interferometer}.
Exactly these internal transitions are the origin of the redshift sensitivity of our proposal: the atom accumulates phase differences caused by the same proper-time difference between the branches, while being sequentially in different internal states.
Hence, our work represents a concept which goes beyond previous ideas of quantum-clock interferometry~\cite{RouraRedShift,Zych1}.

\section{Relativistic description}
To prove the redshift sensitivity, we calculate the phase of such a scheme to first order in $\Delta m=m_\excited-m_\ground$ and $1/c^2$.
The proper time experienced by a particle traveling along a trajectory $z$ with a velocity $\dot{z}$ in a uniform Newtonian gravitational potential $g z$ reads
\begin{equation}
\label{eq:properTime}
\tau=\int\!\! \D \tau=\int \!\!\D t\left[1-\dot{z}^2/2c^2+gz/c^2 \right].
\end{equation}
To describe the interaction with the lasers, one has to take the full multi-level structure of the Hamiltonian into account including all internal-state contributions to the diffraction process.
However, after adiabatic elimination and assuming infinitely short laser pulses, the interferometer sequence reduces to a branch-dependent description for which the Hamiltonian corresponding to the upper branch ($\alpha=u$) and lower branch ($\alpha=l$)
\begin{equation}
\label{eq:Hamiltonian}
\hat{H}^{(\alpha)}=m_\state c^2+\frac{\hat{p}^2}{2m_\state }+m_\state  g \hat{z}+\hat{V}^{(\alpha)}_\mathrm{em}(t,\hat{z})
\end{equation}
contains only relativistic contributions in form of different masses $m_\state$~\cite{Zych1, Sonnleitner, Schwartz}.
Laser pulses change the internal state and consequently the mass, turning it into a dynamical quantity.

The validity of \Eq{eq:Hamiltonian} relies on a differential scheme using inverted internal states in subsequent shots, i.\,e. $m_\excited\leftrightarrow m_\ground$.
Consequently, $1/c^2$ corrections to the Hamiltonian either drop out in the differential phase if mass independent, or enter proportional to $\Delta m/c^2$, which is beyond the considered order. 
This procedure results in a strong immunity against various relativistic contributions such as corrections to the center-of-mass motion, finite-speed effects of light as well as higher-order Doppler shifts in the momentum recoil.
\begin{figure}
	\begin{center}
		\includegraphics{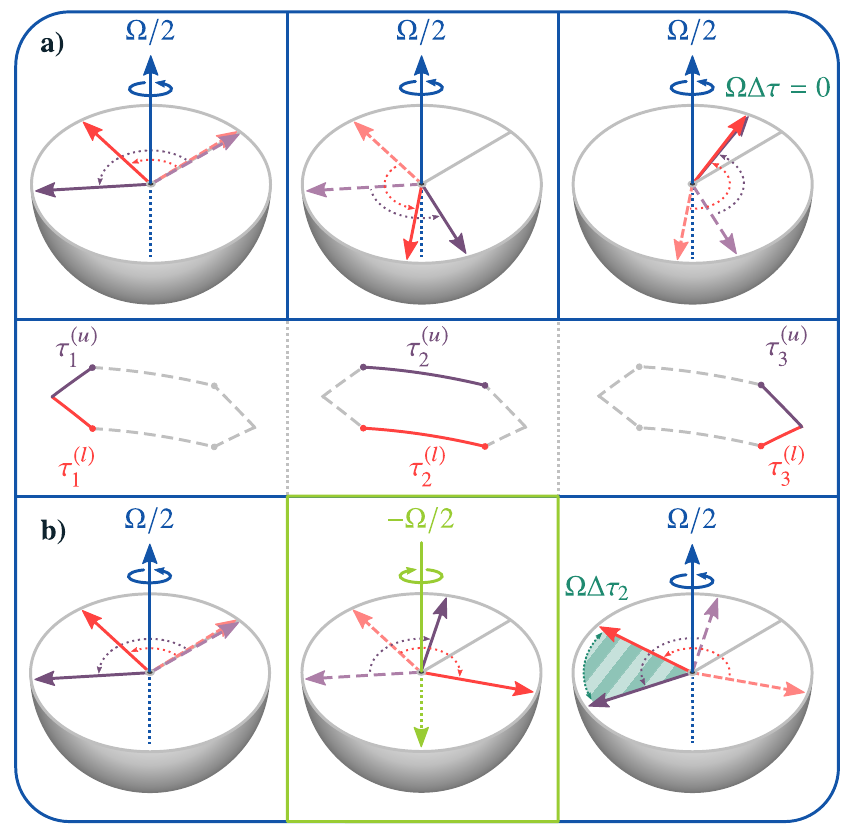}
		\caption{%
		\textit{Origin of the clock phase.} We compare clock phases accumulated between the upper branch (violet arrow) and lower branch (red arrow) along the three segments of the interferometer.
		The accumulation of phases on each branch is depicted by a rotation on the equatorial plane, where the Compton frequency $m c^2/\hbar$ is chosen as a reference. Hence, atoms in the excited state rotate counterclockwise with $\Omega/2$ around the axial vector \myMarker{cset-aps-blue}, whereas the axial vector flips when the atoms are in the ground state \myMarker{cset-aps-limegreen} so that they rotate clockwise with $-\Omega /2$.
		a) If the internal state is not changed during the central segment, the sum of all proper-time differences vanishes so that the clock phase $\Omega \Delta \tau=0$.
		b) If the internal state flips during the central segment, the phase difference does not vanish although $\Delta \tau =0$. Instead, the phases accumulated during each segment add, and one finds the clock phase  $\Omega \Delta \tau_2$, represented by the shaded area. Because the atoms fall during the central segment in parallel at different heights, the phase is proportional to the gravitational redshift.
		} 
		\label{fig:illustrationProperTime}
	\end{center}
\end{figure}

The laser-atom interaction is described by an effective potential~\cite{RedshiftSchleich} of the form $\hat{V}_{\mathrm{em}}^\branch=-\hbar \sum_\ell[k_\ell^\branch\hat{z}+\varphi_\ell^\branch]\delta(t-t_\ell)$,
which transfers momentum $\hbar k_\ell^\branch$ to branch $\branch$ at $t=t_\ell$ and imprints on it the laser phase $\varphi_\ell^\branch$ evaluated at the time of the pulse. We omit in the following the trivial contribution from the latter.
Since in our geometry the atoms are always in the same internal state at equal times (in the laboratory frame) so that $m_\state$ is branch independent, we restrict the discussion to such schemes, but the treatment can be generalized if necessary.

Defining the proper-time differences $\Delta \tau_n=\tau_n^{(u)}-\tau_n^{(l)}$ for each segment $n=1,2,3$ associated with one internal state, we find by Taylor expansion of Hamiltonian~\eqref{eq:Hamiltonian} to first order in $\Delta m$ and subsequent application of the perturbative method developed in the Refs.~\cite{UfrechtPerturbation, Ufrecht}, the phase
 \begin{equation}
\label{eq:phaseExpression}
\Delta\phi=\Delta\phi_m-\frac{\Omega}{2}\sum_n\lambda_\state\Delta \tau_n
\end{equation}
where $\lambda_\state=\pm 1$ indicates the internal state. For more details on the derivation of the phase we refer to the Appendix where we also discuss the origin of the phase in the freely-falling frame.
Equation (\ref{eq:phaseExpression}) underlines that, to first order in $\Delta m$ and $1/c^2$, the total phase is the sum of two contributions:
(i.) The reference phase $\Delta\phi_m$ is independent of $\Delta m$ and is obtained with the reference Hamiltonian $\hat{H}_m$ given by \Eq{eq:Hamiltonian} evaluated at the reference mass $m_\state=m$ where $m=(m_++m_-)/2$ and we assumed that the interferometer with respect to $\hat{H}_m$ is closed.
(ii.) A clock phase as a linear combination of proper-time differences of each segment calculated for the trajectories generated by $\hat{H}_m$ (not the total Hamiltonian $\hat{H}$).
Since the proper-time difference for each interferometer segment enters proportional to $\Omega$, the clock phases can be associated with the ticking rate of an atomic clock even \textit{without} an internal superposition.

\section{Redshift-sensitive geometry}
For the proposed interferometer scheme we obtain with the help of \Eq{eq:phaseExpression} the phase $\Delta\phi=\Delta\phi_m- \Omega \left[\Delta \tau_1-\Delta \tau_2+\Delta \tau_3   \right] /2$, where we used $\lambda_\state$ according to the masses shown in \Fig{fig:Interferometer}.
Like in a Mach-Zehnder interferometer, the symmetry leads to a vanishing proper-time difference between the branches (generated by $\hat{H}_m$) so that $ \Delta \tau=\Delta \tau_1+\Delta \tau_2+\Delta \tau_3=0$.
This constraint eliminates the dependence on $\Delta \tau_1$ and $\Delta\tau_3$ and we obtain
\begin{equation}
\label{eq:FinalPhase}
\Delta\phi=\Delta\phi_m+\Omega \Delta \tau_2\,.
\end{equation}
Our result is explained in more detail in \Fig{fig:illustrationProperTime}.
Since during the central segment the velocities of the atoms on both branches are the same, special-relativistic contributions to proper time cancel and the result calculated with the help of \Eq{eq:properTime} depends solely on the gravitational time-dilation factor
\begin{equation}
\Delta \tau_2 =\frac{g \Delta z}{c^2}T=g\frac{ 2\hbar k T_1 T}{m c^2}
\end{equation}
which contains the wave number $k$ since the height difference $\Delta z=2\hbar k T_1/m$ is caused by the separation of the branches after the first laser pulse.

The clock phase $\Omega \Delta \tau_2$ in \Eq{eq:FinalPhase} exactly matches the one measured by two stationary clocks separated by a distance $\Delta z$ in a uniform gravitational field.
Using for example a general formula \cite{TwinPaper} to obtain the reference phase, we furthermore find $\Delta\phi_m=-2 k g T_1(T+T_1)$.
Note that this phase can be used for tests of UFF, however, not for UGR.

\subsection{Violation model}
\label{Violation model}
To see whether our scheme is sensitive to violations of UGR~\cite{RouraRedShift}, we rely on a consistent parametrization of a violation model e.g.~a dilaton field coupled to the Standard Model and general relativity~\cite{Damour2010,Damour2012}.
In the low-energy and non-relativistic limit the substitution~\cite{RedshiftDebate2} $g\rightarrow(1+\beta_\state)g$ characterizes the impact of this massless scalar particle on the phase in \Eq{eq:FinalPhase}.
Hence, the coupling of the test mass to gravity is no longer universal but occurs in a state-dependent manner represented by the parameter $\beta_\state$, associated with the mass $m_\state$ and this coupling mirrors possible violations of both UFF and UGR. Consequently, the interferometer phase $\Delta\phi$ from \Eq{eq:FinalPhase} is modified
\begin{equation}
\label{eq:DilatonPhaseBlanc}
\Delta\phi=\Omega\Delta\tau_2\left(1+\alpha\right)-2kgT_1\left(T+T_1\right)\left(1+\beta_\excited\right)
\end{equation}
to include
\begin{equation}
\label{ConnectionParameters}
\Delta m\,\alpha=m\left(\beta_\excited-\beta_\ground\right)\,,
\end{equation}
where we neglected terms of the form $\Delta m \beta_\state$. 
For reasonable experimental parameters (see Sec.~\ref{Experimental realization}) the reference phase $\Delta \phi_m = -2 k g T_1 (T+T_1)$ can be larger than $5\cdot10^8$ rad while the clock phase $\Omega \Delta \tau_2$ is about eleven orders of magnitude smaller at the level of a few milliradians. 
Consequently, read-out of the phase, for example through a laser frequency chirp, requires high control and any spurious phases have to be well suppressed.

Our implementation relies on the suppression of common phase shifts in successive state-inverted runs which, to first order, includes the reference phase $\Delta \phi_m$. Here, we assume these phase shifts to be constant within subsequent cycles which will require an experimental validation.
In Sec.~\ref{Experimental realization} we discuss an interleaved operation to reduce the impact of vibration noise which remains in our scheme. An alternative route resulting in high degree of noise suppression is the simultaneous operation of the interferometer with both internal states (either in a mixture or a superposition) in the spirit of quantum clock interferometry. A differential suppression of acceleration noise has recently been proven to a relative uncertainty of  $10^{-12}$  \cite{KasevichWEP}.

Neglecting noise and spurious phase shifts, the differential signal for two successive runs is calculated in leading order from \Eq{eq:DilatonPhaseBlanc} and leads to
\begin{equation}
\label{UGRandUFF}
    \delta \phi=2\Omega\Delta\tau_2\left(1+\alpha\right)-2kgT_1\left(T+T_1\right)\left(\beta_\excited-\beta_\ground\right),
\end{equation}
which offers the possibility to test \emph{both} UGR (first term) and UFF (second term) as detailed in the following. Both contributions can be experimentally distinguished  by varying the duration $T$ of the central segment.
To rigorously associate the different phase contributions to UFF and UGR, we resort to a classical redshift measurement with two independent clocks as first performed by Pound and Rebka \cite{Pound1959}. Introducing for this landmark experiment the same violation parameters as presented above \cite{Will06}, the phase difference between the two clocks takes the form $\Omega g(1+\alpha) \Delta z T/c^2$ \cite{RedshiftTwoClocks}.
It consists of (i) a frequency $\Omega$ multiplied by (ii) the proper-time difference induced by gravity and (iii) a factor of unity modified by the violation parameter $\alpha$. This functional dependence equals \emph{exactly} the first term in \Eq{UGRandUFF} and describes therefore by our definition violations of the universality of the gravitational redshift through $\alpha$.

In contrast, a classical UFF test compares the acceleration of two objects in the gravitational field and violations are characterized by the \emph{Eötvös parameter} determined by the differential acceleration \cite{Eoetvoes}.
Applied to atom interferometric tests with two \textit{different} atomic species \cite{WEP3, KasevichWEP}, one measures a phase difference taking \emph{exactly} the form of the second term in \Eq{UGRandUFF}, where $\beta_\pm$ characterize violations for different atomic species.
Likewise, our proposed scheme compares two internal states of the same atom~\cite{WEPSameSpecies}. In this case the second term in \Eq{UGRandUFF} describes violations of the UFF  through $\Delta \beta =\beta_\excited-\beta_\ground$ where $\beta_\pm$ corresponds to two distinct internal states of the same atomic species. To lowest order, this phase contribution is directly proportional to the Eötvös parameter defining an UFF test.

In most physical parametrizations of theories beyond general relativity the two parameters $\alpha$ and $\Delta \beta$ are interconnected, in case of the dilaton model via \Eq{ConnectionParameters} so that the parameters are proportional to each other.
Even though UFF and UGR might be fundamentally connected,  they  nevertheless  represent  different  principles~\cite{Will06,will_2018}. 
This manifests itself for example by the form of the appearance of the violation parameters as we detail in the following:
By definition UFF tests are Null tests and the second term in \Eq{UGRandUFF} \emph{vanishes} in case no violation is present. In contrast and in complete analogy to classical clocks, UGR tests probe for violations that \emph{alter} the nonzero signal attributable to a proper-time difference in the gravitational potential, so that  $(1+\alpha)$  is determined instead of merely $\alpha$. For this reason we distinguish semantically between the two principles. Consequently, our proposed scheme can be used to set bounds on violation parameters for UFF and UGR in complete analogy to their defining classical counterparts.

\subsection{Experimental realization}
\label{Experimental realization}
\begin{figure}
	\begin{center}
		\includegraphics{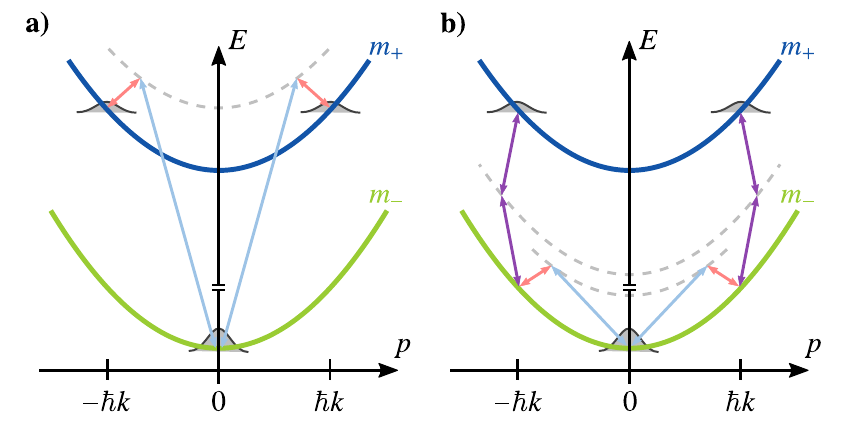}
		\caption{\textit{Energy-momentum diagrams of state- and momentum-changing diffraction schemes.}
		\label{SecDiff} 
		\label{SecA}{a)} Double-Raman diffraction is based on two two-photon processes mediated by a virtual state (dashed line) and generated by two counterpropagating optical lattices (symbolized by the red and blue arrows). Initially in the ground state, denoted by its mass $m_\ground$ \myMarker{cset-aps-limegreen}, the atom is elevated to the excited state with $m_\excited$ \myMarker{cset-aps-blue} and simultaneously diffracted into a superposition of two opposite momenta $\pm \hbar k$. The parabolas highlight the dispersion relation in each state and the condition for a resonant transition.
		\label{SecB}{b)} A sequence of double-Bragg diffraction (red and blue arrows) that first diffracts within the ground state from two counterpropagating lattices into two opposing momentum states in combination with  Doppler-free E1-M1 two-photon transitions (purple arrows) reproduces the double-Raman scheme. However, the recoil-less transitions can be performed with optical frequencies so that a different range of clock frequencies can be addressed compared to conventional double-Raman diffraction.
		}
		\label{fig:Diffraction}
	\end{center}
\end{figure}

A realization requires atomic species with large internal transition frequencies. Recent experiments have demonstrated the coherent control of $^{88}\mathrm{Sr}$~\cite{Strontium} with a transition frequency of the order of a few hundred THz in the optical regime. 
The combination of diffraction with an internal transition naturally occurs for
double-Raman scattering \cite{DoubleRamanInt,HartmannJenewein}, see  \Fig{fig:Diffraction}\,\hyperref[SecA]{\rmfamily a)}, which has also been successfully implemented for gravimetry~\cite{DoubleRamanGravimeter}.
The value of $k$ can be fixed for both experimental runs by chirping the lasers appropriately, which requires four laser frequencies to realize the double-Raman scheme. However, an analog scheme for optical Raman transitions requires laser frequencies in the ultraviolet and has not yet been implemented. 
As a promising alternative to double-Raman diffraction, recent results \cite{Jaffe18} can be generalized so that each pulse is decomposed into a momentum-transfer pulse based on double-Bragg diffraction~\cite{Ahlers16} or a dual-lattice Bloch beam splitter~\cite{BlochBeamsplitter,TwinLattice} and a state-changing pulse employing Doppler-free optical E1-M1 two-photon transitions~\cite{Alden14} accessible for example in $^{88}\mathrm{Sr}$ or $^{170}\mathrm{Yb}$ atoms, see \Fig{fig:Diffraction}\,\hyperref[SecB]{\rmfamily b)}.
Since the momentum transfer must not change upon state inversion, one possibility is to perform double-Bragg diffraction or lattice beam splitters at the magic wavelength~\cite{magwavelength} on the respective states of the clock transition.
Moreover, in both cases one can further increase the separation of the two branches using large-momentum transfer techniques like Bloch oscillations~\cite{Bloch1, FineStructure4, BlochBeamsplitter,TwinLattice}.
In both cases, the lasers address each branch at slightly different times because of the finite speed of light.
Consider a transition from $m_-$ to $m_+$ at time $t_\ell$ in the case of E1-M1 transitions. The phase shift $\Omega \delta t$ due to this delay $\delta t$ is  canceled by the effective laser phase  $\varphi^{(u)}(t_\ell+\delta t)-\varphi^{(l)}(t_\ell)=-\Omega \delta t$ imprinted on the two branches if the laser exactly addresses the resonance. 
In addition, for the same momentum transfer $\hbar k$ in the state-inversed scheme, corrections to the reference phase caused by this non-simultaneity cancels differentially, while the remaining contributions are suppressed by $1/c$ compared to the phase of interest. In case of double-Raman diffraction a similar argument applies.

While the symmetry of our geometry suppresses several other noise sources~\cite{DoubleRamanInt,DoubleRamanGravimeter, Ahlers16}, we expect the signal to be dominated by vibration noise. For a rough estimate of the sensitivity we assume parameters for $^{88}\mathrm{Sr}$, that is  $\Omega=2\pi\cdot  429\,\mathrm{THz}$ and  $k=8\cdot{4\pi}/(813\,\mathrm{nm})$, the latter corresponding to the transition used for magic Bragg and an eight-fold momentum transfer achievable by large-momentum-transfer techniques.  Furthermore, for $T_1=0.25\,\mathrm{s}$ and $T=(0.5\pm0.1)\,\mathrm{s}$ the phase to be resolved is of the order of mrad.
Moreover,  a flux of $10^5$ atoms per second and a vibration-noise equivalent of $5\cdot10^{-10}\,\mathrm{m/s^2}$ is assumed.
The latter is significantly below the performance of current atomic gravimeters~\cite{Hu2013PRA}, but may be reached through a combination of high-performance vibration isolation and classical sensors for post correction, whereas the other parameters are in reach with the state of the art \cite{Gradiometrie4, FineStructure4, BlochBeamsplitter,Bloch1,Compensation2,TwinLattice}.
Alternating the internal states and $T$ between 0.4\,s and 0.6\,s implies a cycle time of 4\,s for one measurement of $\alpha$ and $\Delta\beta=\beta_+-\beta_-$, assuming a preparation of atoms simultaneous to the previous interferometer sequence.
Under these conditions we find after a single cycle a shot-noise limited measurement for $\alpha$ of $4$ and for $\Delta\beta$ of $10^{-10}$ strongly exceeded by vibration noise.
Operating the experiment in a continuous mode with shared pulses~\cite{Savoie2018ScAdv} leads to a suppression of phase noise by $\sim1/t_{\mathrm{avg}}$ when averaging the signal for a time $t_{\mathrm{avg}}$, whereas shot noise decreases as $\sim1/\sqrt{t_{\mathrm{avg}}}$.
Consequently, the noise in $\alpha$ is about 0.03 and in $\Delta\beta$ about $9\cdot10^{-13}$ after $t_{\mathrm{avg}}=6\cdot10^{4}\,\mathrm{s}$.
This sensitivity is three orders of magnitude lower than in state-of-the-art classical redshift tests~\cite{RedshiftTwoClocks}. However, these experiments are performed at different length and time scales so far inaccessible to atom interferometers.
Instead of the interleaved operation, the interferometer could be also performed for both internal states simultaneously in the spirit of quantum clock interferometry to further suppress vibration and laser phase noise.
Such a realization requires Bragg diffraction at the magical wavelength simultaneously for both states. 

The dependence of the phase on initial conditions of the wave packet which might differ from shot to shot due to gravity gradients and rotations is assumed to be successfully mitigated~\cite{Compensation1,Compensation2, Compensation3}.
The approximation  of infinitely short laser pulses is justified for pulse times $\Delta t$ much smaller than the total interferometer time $T$~\cite{Bertoldi}. Due to the differential measurement, finite-pulse-time effects are suppressed by $\Delta t/T\ll1$ compared to the phase of interest.

\section{Discussion}
Since the phase difference of two atomic clocks separated by $\Delta z$ takes the form $\Omega T(1+\alpha) g \Delta z / c^2 $ that also arises in our geometry, classical tests of UGR display in principle the same sensitivity for similar parameters $\Omega $, $T$ and  $\Delta z$, neglecting technical noise.
Therefore, atom interferometers are intrinsically limited by the dimensions of the apparatus, namely  $T$ and $\Delta z$, whereas there is in principle no bound on the separation of two independent clocks.
However, the conceptional relevance of atom interferometric tests of UGR relies on the use of single, delocalized  quantum objects instead of two independent clocks as used for conventional tests.
Hence, our geometry allows probing light-matter coupling and relativistic effects with a delocalized particle, whereas clocks probe the laser field only at localized and independent points in spacetime. Additionally, light-pulse-based interferometers with atoms in free fall offer a complementary approach to atomic clocks since effects such as limited coherence times inside atomic traps due to continuous electromagnetic interactions have not to be actively mitigated~\cite{TrapsClocks}.

\begin{acknowledgments}
E.G., A.F., F.D.P., and C.U. thank Holger Müller for fruitful discussions at the conference ``Quantum Metrology and Physics beyond the Standard Model'' in Hannover.
Furthermore, we thank Sina Loriani and Thomas Hensel for extensive discussions and feedback regarding the presented material.
This work is supported by the German Aerospace Center ({Deutsches Zentrum f\"ur Luft- und Raumfahrt}, DLR) with funds provided by the Federal Ministry for Economic Affairs and Energy ({Bundesministerium f\"ur Wirtschaft und Energie}, BMWi) due to an enactment of the German Bundestag under Grant No. DLR~50WM1556 and No. 50WM1956.
E. G. thanks the German Research Foundation (Deutsche Forschungsgemeinschaft, DFG) for a Mercator Fellowship within CRC 1227 (DQ-mat).
C.~S., D.~S. and E.~M.~R. thank CRC 1227 (DQ-mat) project B07, the EXC 2123 ``Quantum Frontiers'' within the research units B02 and B05, the QUEST-LFS and ``Nieders\"achsisches Vorab'' through ``F\"orderung von Wissenschaft und Technik in Forschung und Lehre'' for the initial funding of research in the new DLR-SI Institute. C.~S. thanks ``Nieders\"achsisches Vorab'' through the ``Quantum- and Nano-Metrology (QUANOMET)'' initiative within the project QT3.
D.~S. gratefully acknowledges funding by the Federal Ministry of Education and Research ({Bundesministerium f\"ur Bildung und Forschung}, BMBF) through the funding program Photonics Research Germany under Contract No. 13N14875.
C.~U., F.~D.~P, A.~F., A.~R.,  E.~G.,  and W.~P.~S. thank the Ministry of Science, Research and Art Baden-Württemberg ({Ministerium f\"ur Wissenschaft, Forschung und Kunst Baden-W\"urttemberg}) for financially supporting the work of IQ$^\mathrm{ST}$.
Moreover, W.~P.~S. is grateful to {Texas A\&M University} for a Faculty Fellowship at the Hagler Institute for Advanced Study as well as to Texas A\&M AgriLife for its support.
\end{acknowledgments}

\section*{Appendix}

\appendix

\section{Calculation of the phase in the laboratory frame}
\label{Calculation of the phase in the laboratory frame}
In order to calculate the phase of the interferometer scheme proposed in the main article, we resort to the perturbative method developed in Refs.~\cite{Ufrecht, UfrechtPerturbation}. The branch-dependent Hamiltonian ($\alpha=u,l$)
\begin{equation}
\label{eq:HamiltonianA}
\hat{H}^{(\alpha)}=m_\state c^2+\frac{\hat{p}^2}{2m_\state }+m_\state  g \hat{z}+\hat{V}^{(\alpha)}_\mathrm{em}(t,\hat{z})
\end{equation}
governing the evolution through the interferometer contains relativistic effects in form of the different masses $m_\state$ corresponding to the internal states and the mass-independent effective laser-atom interaction potential
\begin{equation} \hat{V}_{\mathrm{em}}^\branch=-\hbar \sum_\ell[k_\ell^\branch\hat{z}+\varphi_\ell^\branch]\delta(t-t_\ell)\,.
\end{equation}

The phase $\Delta \phi$ and contrast $C$ of an interferometer are defined by the expectation value
\begin{equation}
\label{eq:OverlapA}
\langle \hat{U}^{(l)\dagger}\hat{U}^{(u)}\rangle =C\e^{\I \Delta\phi}\,,
\end{equation}
where $\hat{U}^{(l)}$ and $\hat{U}^{(u)}$ are the time-evolution operators  with respect to Hamiltonian (\ref{eq:HamiltonianA}) corresponding to the upper branch  and the lower branch.
To calculate the phase to first order in $\Delta m=m_\excited-m_\ground$, we Taylor-expand \Eq{eq:HamiltonianA} around $m_\state=m+\lambda_\state \Delta m/2$ where $m=(m_\ground+m_\excited)/2$ is the reference mass and the dynamical variable $\lambda_\state$ with $\lambda_\ground=-1$ and $\lambda_\excited=+1$ indicates the internal state.
This approximation allows for the decomposition $\hat{H}^\branch=\hat{H}_m^\branch+\hat{H}_{\Delta m}^\branch$ with
\begin{equation}
\label{eq:OperatorProperTimeA}
\hat{H}_{\Delta m}^\branch=\lambda_\state\frac{\Delta m c^2}{2}[1-\left(\hat{p}/m\right)^2/2c^2+g \hat{z}/c^2]
\end{equation}
and $\hat{H}_m^\branch$ is  \Eq{eq:HamiltonianA} evaluated at $m_\state=m$.
We now assume that for  the reference Hamiltonian $\hat{H}_m^\branch$, as in the case of our scheme, the interferometer is closed 
leading to a perfect wave-packet overlap at the end of the interferometer sequence.
Defining $\hat{U}_{m}^\branch$ as the time-evolution operator with respect to $\hat{H}_m^\branch$, this fact translates into $\hat{U}_{m}^{(l)\dagger}\hat{U}_{m}^{(u)}=\mathrm{e}^{\mathrm{i}\Delta \phi_m}$, where $\Delta \phi_m$ is merely a $c$-number. 
When we transform \Eq{eq:OverlapA} into the interaction picture with respect to $\hat{H}_m^\branch$, we find
\begin{align}
\begin{split}
\hat{U}^{(l)\dagger}\hat{U}^{(u)}&=\hat{U}_{\Delta m,I}^{(l)\dagger}\hat{U}_{m}^{(l)\dagger}\hat{U}_{m}^{(u)}\hat{U}_{\Delta m,I}^{(u)}\\
\label{eq:transfomrationA}
&=\e^{\I\Delta\phi_m}\hat{U}_{\Delta m,I}^{(l)\dagger}\hat{U}_{\Delta m,I}^{(u)}\,.
\end{split}
\end{align}
Here, $\hat{U}_{\Delta m,I}^\branch$ is the time-evolution operator associated with $\hat{H}_{\Delta m, I}^\branch$ where the subscript $I$  denotes the interaction picture with respect to $\hat{H}_m^\branch$.
The transformation into the interaction picture amounts to replacing the momentum and position operators in \Eq{eq:OperatorProperTimeA} by their respective solution of the Heisenberg equation of motion generated by $\hat{H}_m^\branch$.
They take the form $\hat{p}(t)=\hat{p}+p(t)$ and $\hat{z}(t)=\hat{z}+\hat{p}t/m +z(t)$, where $p(t)$ and $z(t)$ (without hat) denote the classical trajectories.
To combine the two time-evolution operators on the right-hand side of \Eq{eq:transfomrationA}, we disregard the time-ordering consistently to first order in $\Delta m$ with the help of the Magnus expansion ~\cite{UfrechtPerturbation, Ufrecht}.
To the same order we furthermore merge the two exponents.
Since all wave-packet effects are common to both branches, they cancel and the interferometer is closed.

Defining the proper-time differences $\Delta \tau_n=\tau_n^{(u)}-\tau_n^{(l)}$ for each segment associated with one internal state, we find the phase
 \begin{equation}
 \label{eq:phaseExpressionA}
\Delta\phi=\Delta\phi_m-\frac{\Omega}{2}\sum_n\lambda_\state\Delta \tau_n
\end{equation}
with the help of $\dot{z}(t)=p(t)/m$ and the expression for the proper time to order $1/c^2$
\begin{equation}
\label{eq:properTimeA}
\tau=\int\!\! \D \tau=\int \!\!\D t\left[1-\dot{z}^2/2c^2+gz/c^2 \right].
\end{equation}
We furthermore introduced the transition frequency $\Omega=\Delta m c^2/\hbar$.

\section{Freely-falling frame}
\label{Freely-falling frame}
To understand the origin of the phases, we analyze the interferometer in the freely-falling frame.
From the perspective of a freely-falling observer the trajectories are straight lines.
In our model we assume that the laser pulses act simultaneously on both branches in the laboratory frame since corrections from the finite speed of light are experimentally negligible as described in Sec.~\ref{Experimental realization}.
In the freely-falling frame, however, the laser pulses, acting at time $t$ in the laboratory frame, address each branch at slightly different times $t^\prime$.
The different times $t$ and $t^\prime$ are connected by the Rindler transformation 
\begin{equation}
\label{eq:TimeTransformation}
t^\prime=t\left[1+g^2 t^2/(6c^2)+g z(t)/c^2\right]\,,
\end{equation}
which depends on the position of the atoms in the laboratory frame.
Consequently, the proper-time difference between the two branches of the central segment is still given by $\Delta \tau_2=g\Delta z T/c^2$ in the freely-falling frame.
\begin{figure}
	\begin{center}
		\includegraphics{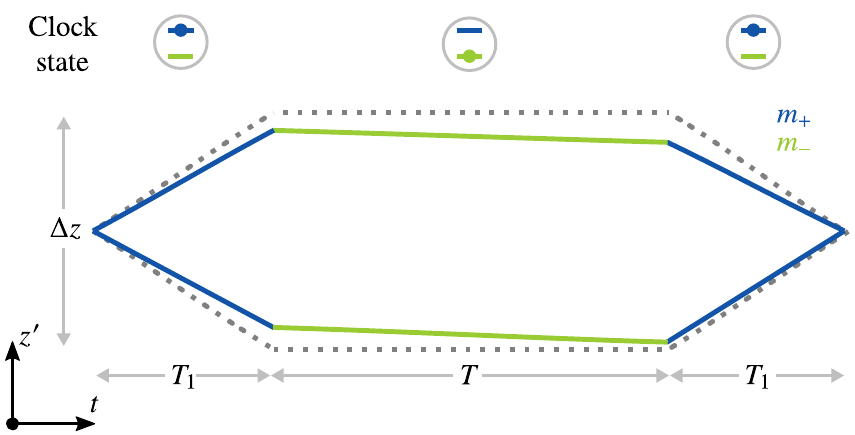}
		\caption{\textbf{Origin of the phase in the freely-falling frame}. 
			Because of the different masses involved, the classical trajectories (solid lines) deviate from those corresponding to $m_\state=m$ (dashed lines). In particular, at time $T_1$ where the laser pulses change the mass from $m_\excited$ \myMarker{cset-aps-blue}  to $m_\ground$ \myMarker{cset-aps-limegreen},
			  momentum conservation in the laboratory frame implies  $p_\mathrm{f}= p_\mathrm{i}-\hbar k$ where $p_\mathrm{i}$   ($p_\mathrm{f}$) is the momentum before (after) the laser pulse. Performing the Galilei transformation, we find $p^\prime_\mathrm{f}-m_\ground gT_1= p^\prime_\mathrm{i}-m_\excited gT_1-\hbar k$.  Consequently, we observe the additional recoil $\Delta p^\prime=-\Delta m g T_1$ in the freely-falling frame resulting in a small  gravity-dependent velocity, which results in  slightly inclined solid lines in the middle segment. It is this residual recoil which introduces the coupling between $\Omega$ and $g$ in the freely-falling frame.}
		\label{fig:FreelyFallingInterferometer}
	\end{center}
\end{figure}
However, within the framework of Hamiltonian (\ref{eq:Hamiltonian}), it is sufficient to perform a Galilei transformation, that is $z^\prime=z+1/2gt^2$ since higher corrections are suppressed by at least the factor $\Delta m/c$ in differential measurements.
As explained in  the caption of \Fig{fig:FreelyFallingInterferometer}, where we plot the transformed trajectories against the laboratory time, it is the deviation of the trajectories due to the change of mass in the presence of the laser pulses which introduces the coupling between $\Omega$ and $g$.

To calculate the phase from the perspective of a freely-falling observer, we transform  Hamiltonian (\ref{eq:Hamiltonian}) into the freely-falling frame, for example by returning to the classical Lagrangian picture. After performing the Galilei transformation and partial integration, we obtain the quantized Hamiltonian
\begin{equation}
\label{eq:HamiltonianFreelyFalling}
\hat{H}^{\prime (\alpha)}=m_\state c^2+\frac{\hat{p}^{\prime 2}}{2m_\state}+\hat{V}^{\prime\branch}_{\mathrm{em}}-m_\state g^2t^2-\dot{m}_\state \hat{z}^\prime g t\,.
\end{equation}
Here, $\dot{m}_\state$ is the time derivative of the mass which is only non-vanishing during the laser pulses.
The  transformed laser-atom interaction takes the form $\hat{V}^{\prime\branch}_{\mathrm{em}}= -\hbar \sum_\ell[k_\ell^{(\alpha)}\hat{z}^\prime+\varphi_\ell^{\prime\branch}]\delta(t-t_\ell)$, where $\varphi_\ell^{\prime\branch}=\varphi_\ell^\branch-1/2gt^2$ is imprinted on the wave packet at each laser pulse.
We again decompose the Hamiltonian through a Taylor expansion into $\hat{H}^{\prime\branch}=\hat{H}_m^{\prime\branch}+\hat{H}_{\Delta m}^{\prime\branch}$ with $\hat{H}_m^{\prime\branch}=\hat{p}^{\prime2}/2m+\hat{V}^{\prime\branch}_{\mathrm{em}}$ and
a Hamiltonian taking into account the changes of the mass
\begin{equation}
\label{eq:OperatorProperTimeFreelyFalling}
\hat{H}_{\Delta m}^{\prime\branch}=\lambda_\state\frac{\Delta m c^2}{2}\left[1-\frac{\left(\hat{p}^\prime/m\right)^2}{2c^2}-\frac{g^2t^2}{c^2}\right]-\dot{m}_\state \hat{z}^\prime g t\,\,
\end{equation}
to first order in $\Delta m$, where we discarded global phases.
By following the derivation of \Eq{eq:phaseExpression}, we find, after recalling $\dot{z}^\prime(t)=p^\prime(t)/m$ and partial integration of the last term in \Eq{eq:OperatorProperTimeFreelyFalling} the phase
\begin{equation}
\label{eq:InvarianceofPhase}
\Delta\phi^\prime=\Delta\phi_m^\prime-\frac{\Omega}{2}\sum_n\lambda_\state \Delta \tau_n^\prime
\end{equation}
in the freely falling frame. To first order in $1/c^2$ the
 proper-time difference  $\Delta \tau_n^\prime= \tau_n^{\prime (u)}- \tau_n^{\prime (l)}$ between the two branches for segment $n$ is defined via the integral
\begin{align}
\tau^\prime=\int_{\mathrlap{}}\mathrm{d}t\frac{\mathrm{d}t^\prime}{\mathrm{d}t}\left[1-\frac{\dot{z}^{\prime2}}{2c^2}\right]
\end{align}
where the Rindler time,  which was defined in \Eq{eq:TimeTransformation}, was identified.
In the freely-falling frame the reference phase $\Delta \phi_m^\prime$ originates 
from the Doppler-shifted laser phase $\varphi_\ell^{\prime\branch}=\varphi_\ell^\branch-1/2gt^2$, which is imprinted on the wave packet at each laser pulse so that $\Delta\phi_m$ takes the same value in both frames.

The invariance of the clock phase simply follows from the invariance of the proper time under coordinate transformations and we find $\Delta \phi=\Delta \phi^\prime$.
Hence, the clock phase indeed can be interpreted as the sum of proper times along the interferometer branches and it does not arise from the Doppler shift of the frequency or wave vector of the laser.

\end{document}